\documentclass[preprint,preprintnumbers,amsmath,amssymb,nofootinbib]{revtex4}
\usepackage{graphicx}
\usepackage{subfigure}
\usepackage{dcolumn}
\usepackage{bm}
\usepackage{enumerate}
\def \prd{{\it Phys. Rev. D, }}

\def \apj{{\it Ap. J., }}

\begin{document}
\title{A Geostationary Gravitational Wave Interferometer (GEOGRAWI)}

\author{Massimo Tinto} \email{Massimo.Tinto@jpl.nasa.gov}
\affiliation{Jet Propulsion Laboratory, \\
  California Institute of Technology, \\
  Pasadena, CA 91109} \altaffiliation [Also at (01/01/ - 12/31, 2012):
]{Instituto Nacional de Pesquisas Espaciais, Astrophysics Division,
  Avenida dos Astronautas 1758, 12227-010 - S\~{a}o Jos\'{e} dos
  Campos, SP, Brazil}

\author{J.C.N. de Araujo}
\email{jcarlos.dearaujo@inpe.br}
\affiliation{Instituto Nacional de Pesquisas Espaciais, Astrophysics
  Division Avenida dos Astronautas 1758, 12227-010 - S\~{a}o Jos\'{e} dos
  Campos, SP, Brazil}
\author{Odylio D. Aguiar}
\email{odylio.aguiar@inpe.br}
\affiliation{Instituto Nacional de Pesquisas Espaciais, Astrophysics
  Division Avenida dos Astronautas 1758, 12227-010 - S\~{a}o Jos\'{e} dos
  Campos, SP, Brazil}
\author{M\'arcio Eduardo da Silva Alves}
\email{alvesmes@unifei.edu.br}
\affiliation{Instituto de Ci\^ encias
  Exatas, Universidade Federal de Itajub\'a, 37500-903 Itajub\'a, MG,
  Brazil}

\date{\today}

\begin{abstract}
  
  We propose a Geostationary Gravitational Wave Interferometer
  (GEOGRAWI) mission concept for making observations in the sub-Hertz
  band.  GEOGRAWI is expected to meet some of LISA's science goals in
  the lower part of its accessible frequency band ($10^{-4} - 2 \times
  10^{-2}$ Hz), and to outperform them by a large margin in the
  higher-part of it ($2 \times 10^{-2} - 10$ Hz). As a consequence of
  its Earth-bound orbit, GEOGRAWI is significantly less expensive than
  the interplanetary LISA mission and could be either an entirely US
  mission or managed and operated by NASA in partnership with the
  Brazilian Space Agency.

\end{abstract}

\maketitle

\section{Objectives and Goals}
\label{intro}

The primary objective of this document, written in response to the
NASA Request For Information \#  NNH11ZDA019L, is to sketch a
mission concept for a space-based detector of gravitational radiation
capable of meeting most of LISA's \cite{PPA98} science goals at a
significantly reduced cost. This mission could be either entirely
managed and operated by NASA or flown in partnership with the
Brazilian Space Agency. In this case the spacecraft busses and the
onboard and ground telecommunication hardware could be provided by the
Brazilian Space Agency, further reducing NASA's mission costs.

\section{Mission Design and Orbit}
\label{Mission}

Our proposed space-based detector entails three spacecraft in
geostationary orbit, forming an equilateral triangle with armlength of
about $73,000$ km. The main advantage of such an interferometer over
LISA is that it is significantly less expensive to launch and position
it in its final orbit. Because of its smaller and constant armlength,
further instrument simplifications over that baselined for LISA follow
as additional benefits.  For instance, no laser ranging modulations
nor modulations needed by the Ultra-Stable Oscillator noise
cancellation scheme will be required; no articulation of the optical
telescopes onboard each spacecraft will need to be implemented; the
attitude control subsystem and onboard propulsion units will be
down-scaled accordingly to the less stringent needs imposed by the
spacecraft trajectories; ground data acquisition can be performed with
three small dedicated antennas whose cost is a fraction of the tracking
costs LISA would require; in the eventuality of system/subsystem
failure a robotic repair mission could be performed. This list of
advantages associated with a geostationary configuration is of course
not exhaustive.  However, if we just consider the cost savings
associated with a smaller launch vehicle and propulsion module than
those required by LISA as well as those resulting from the more benign
orbit, we may already see that the cost of a geostationary
interferometer will be smaller than that for LISA. Furthermore, it has
been pointed out in recent years that a single, spherical
proof-mass could be adopted for achieving the desired inertial
reference frame stability required by LISA \cite{Stanford2005}. In our
cost estimate we will assume each spacecraft to be equipped with a
single, spherical proof-mass.

\section{Sensitivities}
\label{SENSITIVITY}

Because the armlength of GEOGRAWI is roughly $70$ times smaller than
the armlength of the LISA mission, its sensitivity to gravitational
radiation in the lower part of its accessible frequency band will be
proportionally worse than that of LISA (assuming of course onboard
accelerometers of similar performance as those baselined for LISA). On
the other end, the shorter baseline of GEOGRAWI will result in a much
smaller photo-counting error at the photo-detectors and an improved
sensitivity over that of LISA by the same factor $70$ in the remaining
(higher) part of its frequency band.

Using the standard definition of sensitivity for a space-based
interferometer \cite{TA1998,AET1999,ETA2000,TD2005}, we have estimated the
Time-Delay Interferometric (TDI) sensitivities of a geostationary
interferometer under the following three different on-board subsystem
configurations:

\begin{enumerate}[(I)]
\item The onboard instrumentation and its noise performance is
  equal to that of the LISA mission. We will refer to this
  configuration as the ``Geostationary LISA''.
  
\item The output power of the onboard lasers and the size of the
  optical telescopes are assumed to be equal to that of the LISA
  mission, while the noise performance of the accelerometers is taken
  to be 10 times worse than that of the accelerometer planned for the
  LISA mission. This configuration will be referred to as the
  ``Geostationary 1''. \footnote{This accelerometer noise level is
    equal to that of the accelerometer to be flown on-board the LISA
    Pathfinder mission \cite{Pathfinder}}

\item The noise performance of each accelerometer is taken to be a
  factor of ten worse than that of the accelerometers planned for the
  LISA mission, the output power of the lasers is assumed to be a
  factor of 10 smaller than that of the lasers onboard LISA, and the
  diameter of the optical telescopes has been reduced by a factor of
  $\sqrt{10}$ over that of the LISA telescopes. This configuration
  will be called ``Geostationary 2''.
\end{enumerate}

\begin{figure}
\centering
\includegraphics[width = 6in]{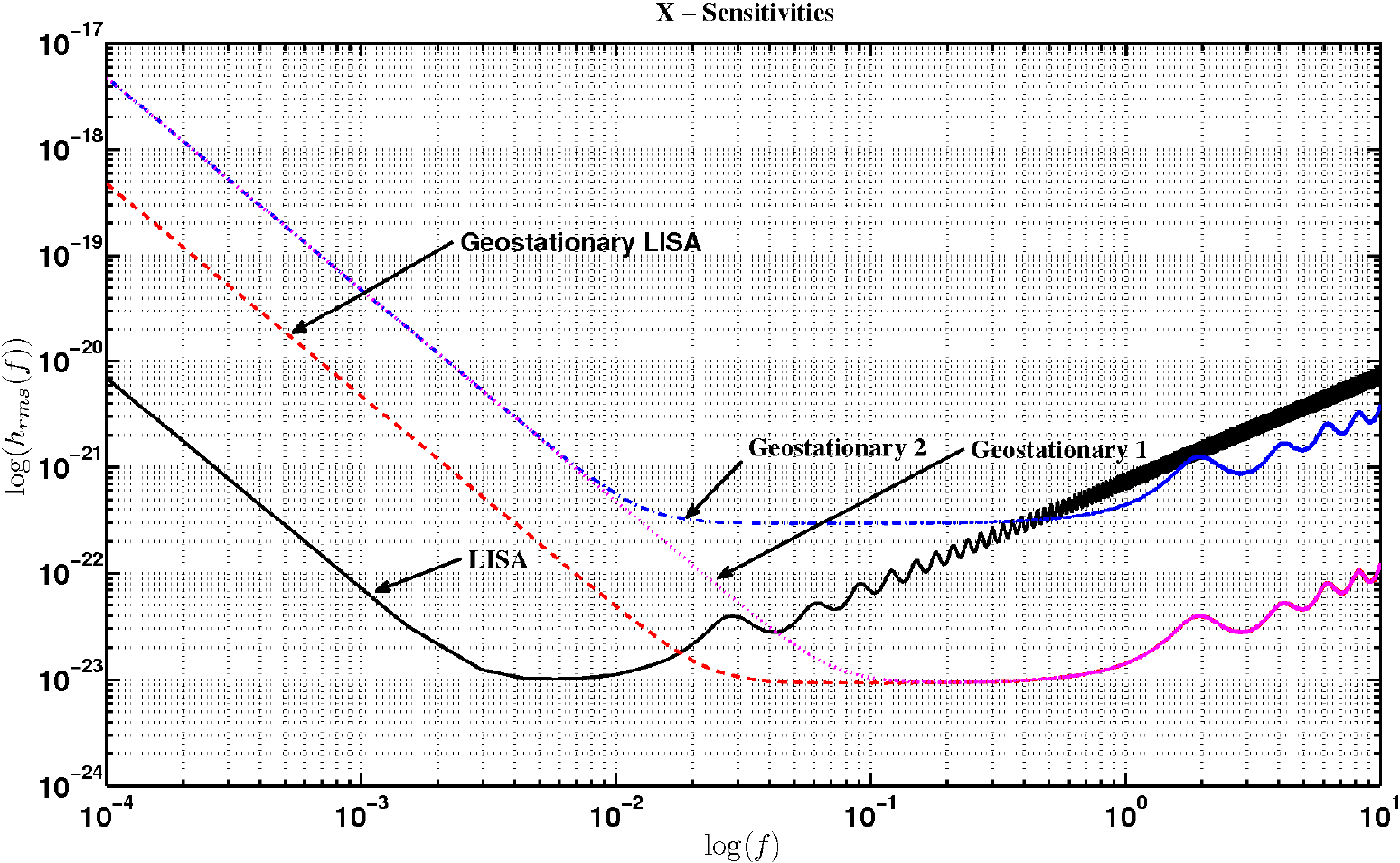}
\caption{Sensitivity of the $X$ combination of a geostationary
  interferometer. Its noise performance is characterized by the noise
  spectra given in \cite{ETA2000} for the LISA mission and properly
  scaled here for the three different onboard subsystem
  configurations. The estimated LISA sensitivity is included for sake
  of comparison.}
\label{Sensitivities}
\end{figure}
In Figure (\ref{Sensitivities}) we plot the sensitivities of the TDI
combination $X$ \cite{TD2005} for the various on-board hardware
configurations discussed above and, for sake of comparison, we include
the LISA sensitivity. At high-frequencies the sensitivity of
any of the geostationary interferometers considered is significantly
better than that of the LISA mission while, at lower frequencies, the
longer armlength of LISA results in a better sensitivity in this
part of the band.

\section{Scientific Objectives}
\label{Science}

The enhanced sensitivity of GEOGRAWI at frequencies larger than about
$20$ mHz implies that it will be able to observe massive and
super-massive Black Holes (SMBHs), stellar-mass binary systems,
several binary systems present in our own galaxy (the so called
``calibrators'') \cite{PPA98}, cosmic strings, and a stochastic
background of astrophysical or cosmological origin. Although GEOGRAWI
will be unable to detect the zero-order cyclic spectrum of the white
dwarf-white dwarf galactic binary confusion noise, it will however
detect and measure the higher-order ``cyclic spectra'' present in the
data because of the rotatory motion of the interferometer around the
Sun.  The cyclic spectra can be observable as they are not affected by
the instrumental noise {\underbar {if}} this is stationary
\cite{Edlund}. Under this assumption, GEOGRAWI will still detect the
so-called ``Confusion Noise'', and infer properties of the
distribution of the white-dwarf binary systems present in our galaxy
\cite{Edlund}.

In relation to binary black-holes systems, we have recently analyzed
how well and how many of them GEOGRAWI will be able to detect, as
these sources were of primary interest to LISA \cite{Araujo2011}.
Since a significant amount of GW energy can be released during the
three evolutionary phases (inspiral, merger and ring-down) of these
systems, we have calculated the maximum redshift, for a given
signal-to-noise ratio (SNR), at which these systems could be
detectable during these three phases.  From these results we then
inferred the event rate by relying on the work recently done by
Filloux {\it et al.} \cite{fill} on the formation and evolution of
massive and supermassive black-holes.  We found that the Geostationary
LISA configuration could see as many as $19$ black-hole binaries per
year with a $SNR = 10$ out to a maximum redshift of $10$. This number
of events rate is slightly larger than that for LISA as a consequence
of the fact that this geostationary interferometer has a better
sensitivity at higher frequencies where smaller black-holes binaries
radiate. Since smaller black-holes are easier to form and are
therefore larger in number, a geostationary LISA will be able to see
more of them than LISA.

\section{Estimated Cost}
\label{Cost}

In order to estimate the cost of a geostationary gravitational wave
interferometer, we have relied on the latest cost estimate for LISA
included in the document titled: ''Laser Interferometer Space Antenna
(LISA) Astro2010 RFI \#2 Space Response'', dated 3 August 2009, which
was submitted by the LISA project to the NRC Decadal Review panel as a
Request For Information \cite{RFILISA}. At page 57 of the above
document a break-down table of the US costs for LISA is provided. We
have used that table to generate a similar breakdown for a
Geostationary LISA, i.e. a geostationary interferometer whose
spacecraft are equipped with the same LISA instrumentation. As
mentioned in the previous section, this is a somewhat conservative
assumption as some instrument simplifications will result from a more
benign mission orbit.

The table below compares the LISA break-down and total costs against
those of a geostationary interferometer detector under the assumptions
of (i) either an entirely US-funded mission, or (ii) a partnership
with the Brazilian Space Agency will lift the NASA costs associated with
the spacecraft, propulsion modules, and ground data system.
Explanations of the main cost differences between a ``Geostationary
LISA'' and LISA are included in the Table. We find the final
NASA cost of a Geostationary LISA to be equal to about \$ 1.1 B
for an entirely US-managed mission. A joint partnership with the
Brazilian Space Agency would further reduce down the NASA costs 
to about \$ 940 M. 

\begin{figure}
\begin{center}
\caption{Cost estimates for a Geostationary interferometer, under the
  assumptions of it being (i) a US-only mission, or (ii) flown in
  partnership with the Brazilian Space Agency. Reserves have been
  assumed to be at the 20 \% level. The cost of the LISA mission is
  included for comparison.}  
\includegraphics[width = 7in]{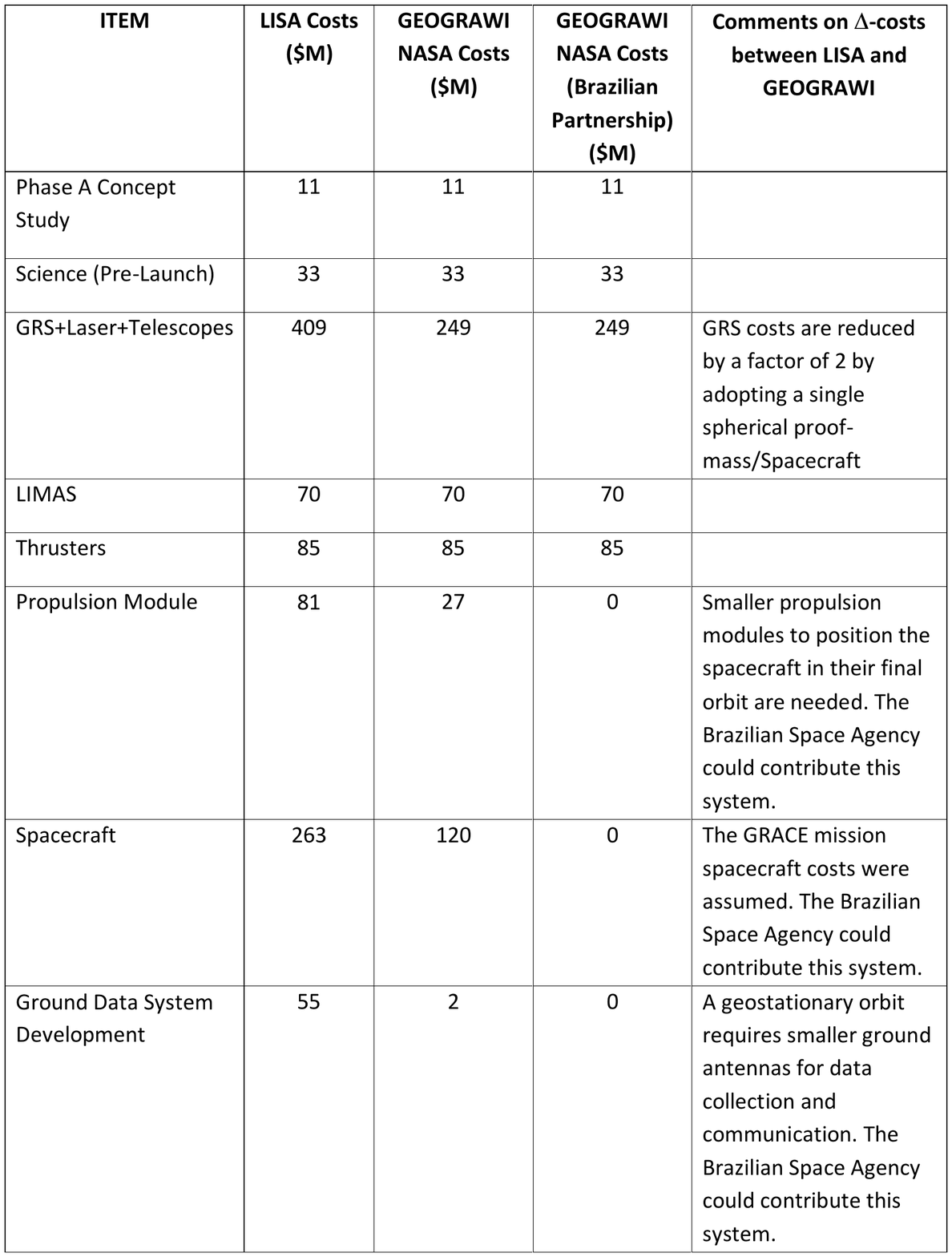}
\end{center}
\label{Costs1}
\end{figure}
\begin{figure}
\begin{center}
\includegraphics[width = 7in]{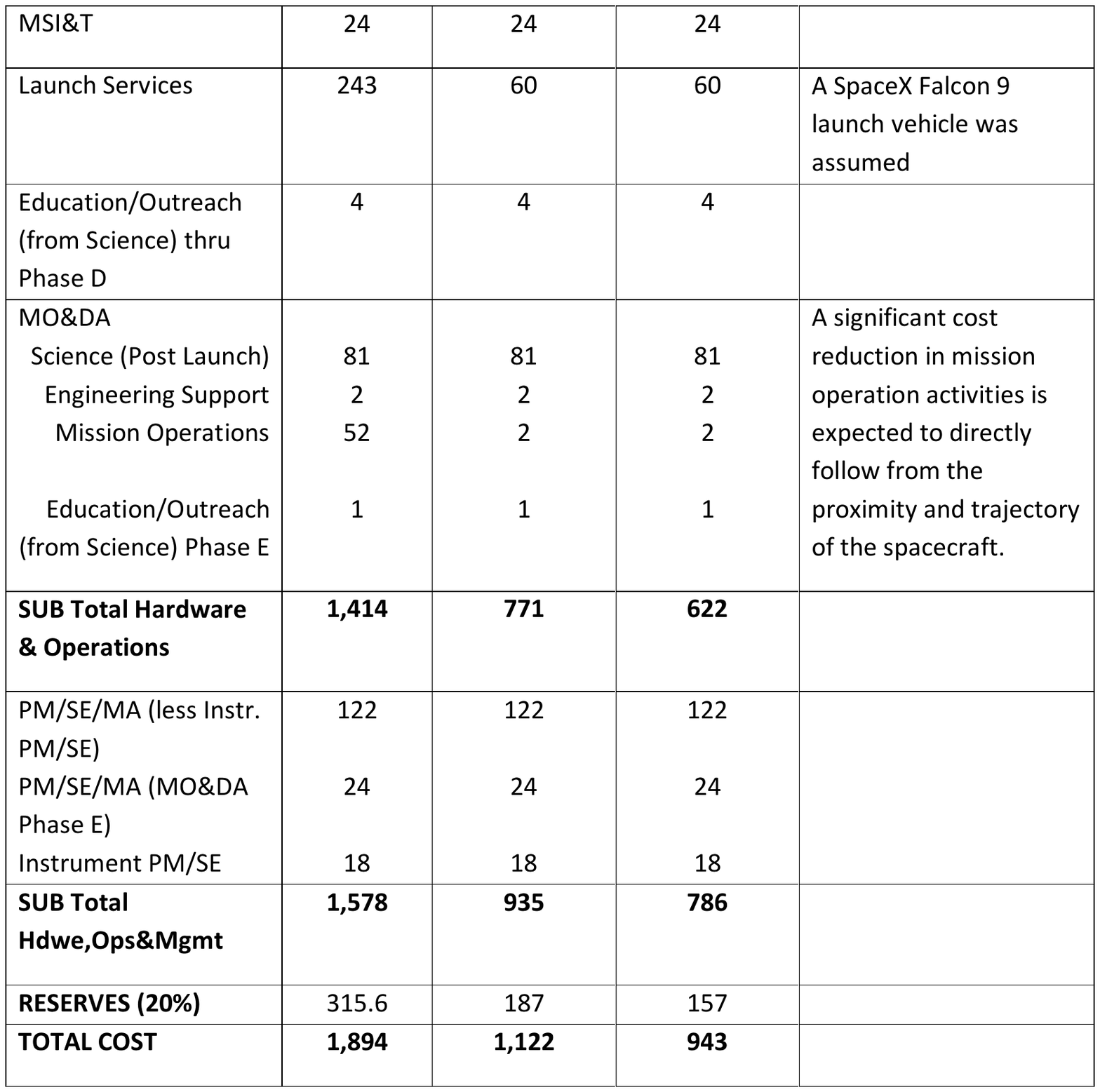}
\end{center}
\label{Costs2}
\end{figure}
\vskip24pt
The cost information contained in this document is of a budgetary and
planning nature and is intended for informational purposes only.  It
does not constitute a commitment on the part of JPL and/or Caltech.

\section{Acknowledgments}

For M.T., this research was carried out at the Jet Propulsion
Laboratory, California Institute of Technology, under a contract with
the National Aeronautics and Space Administration. \copyright \ 2011.
All rights reserved.  J.C.N.A.  and O.D.A. would like to thank CNPq
and FAPESP for financial support.

\eject


\begin{references}
\bibitem{PPA98} P. Bender, K. Danzmann, \& the LISA Study Team, Laser
  Interferometer Space Antenna for the Detection of Gravitational
  Waves, Pre-Phase A Report, MPQ233 (Max-Planck- Instit\"ut f\"ur
  Quantenoptik, Garching), July 1998.
\bibitem{Stanford2005} Ke-Xun Sun, Graham Allen, Sasha Buchman, Dan DeBra
  and Robert Byer, {\it Class. Quantum Grav.}, {\bf 22} S287 - S296,
  (2005)
\bibitem{TA1998} M. Tinto and J.W. Armstrong, \prd {\bf 59}, 102003, (1999).
\bibitem{AET1999} J.W. Armstrong, F.B. Estabrook and M.
Tinto, \apj {\bf 527}, 814 (1999).
\bibitem{ETA2000} F.B. Estabrook, M. Tinto, and J.W. Armstrong,
\prd {\bf 62}, 042002 (2000).
\bibitem{TD2005}M. Tinto and S.V. Dhurandhar, {\it Living Rev.
    Relativity}, {\bf 8}, 4, (2005). http://www.livingreviews.org/lrr-2005-4
\bibitem{Pathfinder} A. Cavalleri {\it et al.},  {\it Class. Quantum
    Grav.}, {\bf 26} 094012 (2009)
\bibitem{Edlund} J.A. Edlund, M. Tinto, A. Kr\'olak, G. Nelemans, {\it
    Classic. Quantum Grav.}, \textbf{22}, S913-S926 (2005)
\bibitem{Araujo2011} J.C. Araujo, O.D. Aguiar, M.E.S. Alves, and M.
  Tinto. In preparation.
\bibitem{fill} Ch. Filloux, J.A. de Freitas Pacheco, F. Durier and
  J.C.N. de Araujo, arXiv:1108.2638 (2011) and IJMPD in press
\bibitem{RFILISA} R.T. Stebbins {\it et al.}, {\it Laser Interferometer Space Antenna (LISA)
Astro2010 RFI \#2 Space Response} (2009) http:\/\/lisa.gsfc.nasa.gov/Documentation/Astro2010\_RFI2\_LISA.pdf
\end{references}
\end{document}